# Synthesis and structural characterization of uranium-doped $Ca_2CuO_3$, a 1D quantum antiferromagnet

Nam Nhat Hoang[1,3], Dang Chinh Huynh[1,4], Duc Tho Nguyen[1],

Thuy Trang Nguyen[1], Duc The Ngo[2], Michael Finnie[2], and Chau Nguyen[1]

[1]Center for Materials Science, College of Science, Vietnam National University, 334 Nguyen Trai, Hanoi, Vietnam

[2] Department of Physics and Astronomy, University of Glasgow, Glasgow G12 8QQ, United Kingdom

**Abstract**

The technological settings of a modified sol-gel method for preparation of highly fine homogeneous powder $Ca_2CuO_3$ doped with uranium 238 ($x$=0–0.05) is presented. The analysis of structure, purity of phases and the justification for the role of uranium in the given compounds are provided together with almost complete classification of observed optical phonons by means of the Raman, IR measurements and *ab initio* calculation. The significant reduction in particle size was achieved by doping and the strong correlation between resistivity and doping concentration was observed and explained using the phonon-assisted electron hopping conduction model. The persistence of covalent insulation state in all compounds is a key feature of this class of compounds.

[3] Author to whom any correspondence should be addressed.

E-mail: namnhat@gmail.com and namnhat@hn.vnn.vn

[4] Also from the Faculty of Chemical Technology, Hanoi University of Technology, 1 Dai Co Viet Str., Hanoi, Vietnam

## I. Introduction

During the last decades scientists have paid much attention to the quasi 1D antiferromagnetic system $Ca_2CuO_3$. As a quantum magnet, this system promises direct application in high-techs and medicine. The compound proved itself as a valuable additive in processing of the Cu-O based high $T_c$ superconductors [1,2]. For the fundamental issues, its close structural relationship to the $La_2CuO_4$-type high $T_c$ superconductors has stimulated the extensive studies to identify the essentials of superconductivity in the low dimension [3-9]. Several doping studies have also been presented, mainly in searching for the new class of the low dimension high $T_c$ superconductors [14-16]. However, there was a lack of studies dealing with the controlling of the covalent insulation state and conductivity of this system, for which various direct applications of the compound depend on. In this paper we present the technological settings of a modified sol-gel method for preparation of the highly homogeneous nanoparticles $Ca_2CuO_3$ doped with uranium 238. We show that the resistivity of the bulk samples (in constancy of its covalent insulation state) can be successfully managed by this doping. The basic importance of this study is of two folds: (1) the pure $Ca_2CuO_3$ has quite high dielectric constant at room temperature and would stimulate the direct application in electronic devices if its conductivity would be well controlled; (2) the $Ca_2CuO_3$ is known to exhibit the femto-second optical excitation life-time so the achievement of the induced optical transition in visible region by doping is a crutial factor for future application of this material in quantum optics. Indeed, a narrow-band transition was observed in all uranium-doped $Ca_2CuO_3$ and this will be discussed in details in a separate paper. A further stimulating factor for this study comes from the difficulties in preparation of the highly homogeneous powder $Ca_2CuO_3$ by means of the ceramic and oxalate co-precipitation techniques [23] so a successful modification of the sol-gel route for the purpose of obtaining the single-phased nanocrystallites would itself be desirable.

Let us briefly introduce the basics of this system. Being crystallized in the space group *Immm* (no.71), the $Ca_2CuO_3$ has the structure (schematically featured in Figure 1(a)) which is very similar to the one of the 2D superconducting $La_2CuO_4$: there is only one oxygen atom lacking which perpendicularly connects two parallel chains of the $CuO_4$ squares. For this reason, along the $CuO_4$ chains the system exhibits a much stronger antiferromagnetic coupling than perpendicular to those chains. The *t-J* model has estimated its intrachain exchange integral $J_{\parallel} \approx 0.6$eV (recordly high among the 1D systems) which is about 300 times greater than the interchain coupling $J_{\perp}$ [3-6,7-9]. Some compounds with the $Ca_2CuO_3$ structure, *e.g.* an oxygen excessive $Sr_2CuO_{3.1}$, can transform its structure under pressure into the $La_2CuO_4$ type structure and become the high $T_c$ superconductor ($T_c$=70K) [10]. Due to low dimensionality, the compound exhibits a series of interesting physical properties, such as covalent insulation [11], van Hove singularity on the spin Fermi surface [37], spin-charge separation [12, 22] *etc*. At room temperature, the solid bulk sample is a covalent insulator with high dielectric constant. Its resistivity depends heavily on preparation routes as there were large differences in reported values [23,30]. The energy gap between the insulation ground state and the conduction band is however very small (as of *meV*). Usually, the transport mechanism in the hole-doped $Ca_2CuO_3$ was regarded as driven mainly by the hopping of the small polarons [16] and as the polaron size could not be exactly controlled, it seemed that there was no way to manage the resistivity of the $Ca_2CuO_3$ within a given region.

For this purpose, we introduce the usage of $U^{238}$ in $Ca_2CuO_3$. In the other ceramics, namely in the PZT (Mg/Nb) pyroelectric ceramics, the insertion of the small amount of uranium (≤1.45%) showed the strong impact on lowering of resistivity by several orders of magnitude [21]. For the Cu-O based high $T_c$ superconductors, the doping of $U^{238}$ has also been introduced and patented (U-doped Nd-Ba-Cu-O in [17-19], U-doped Tl-1223 in [20]). Several problems remained unclear, however, as for the role and the position of uranium in parent structures: some studies reported the uranium resided in isolated islands [17-19], others considered it substituted in the parent lattices [20, 21]. These problems seemed to be associated with the purity of phases (usually prepared by the ceramic technique) which was not fully demonstrated in the above cases.

We show here that the uranium may be doped successfully into the $Ca_2CuO_3$ system via a modified sol-gel route and that the doping structures remain single-phased with the improved refinement of nanoparticle size and substantial change in resistivity (within the constancy of covalent insulation state). The clarification of all optical phonons supports the argument that the doped U is likely to occur at the interstitial regions between the monocrystal surfaces.

## II. Preparation

In the previous paper Ref. [22] we have addressed the efficiency of a modified sol-gel method in the preparation of the single-phased $Ca_2CuO_3$. In the classical solid-state reaction technique, because of the dissociation of CuO into $Cu_2O$ and $O_2$ at high temperature, the deficit of the reacting CuO leads usually to the impure phases $CaCu_2O_3$ and CaO. In case the CuO is excessively added, its actual amount in solid solution is thus non-stoichiometric and the final product often contains large amount of the non-reacting CuO. This problem even appeared serious for the oxalate co-precipitation technique [1] and for the travelling-solvent floating-zone method (for growing of the single crystals, near 5% of the $CaCu_2O_3$ phase has been indicated [24]). As the sol-gel route does not use the precursors CuO and Cao, so the impact of CuO dissociation is reducing, we modified this route for the purpose of preparation of the uranium doped $Ca_2CuO_3$.

Of the following, all chemicals used were at high grade (99.9% or more) and come from Merck and MaTeck. First, the amount of the starting ammonium diuranate $(NH_4)_2U_3O_7$ was chosen for the required stoichiometry $Ca_2CuO_3$:$U_x$, $x$=0.0, 0.005, 0.01, 0.025 and 0.05. This amount was dissolved in water and mixed with the $Ca(NO_3)_2$ and the $Cu(NO_3)_2$ solutions to form the mixture of the metal nitrates at desired cation proportions. The citric acid $C_6H_8O_7$ (CA) as the reagent was added to this mixture so that the proportion of the reagent over all metal ions was fractionally over 3:1. Afterwards the $NH_4OH$ solution was droped in slowly and with vigorous stirring to achieve the pH near 3.5. In continuous stirring, the solution was heated up to $80^0$C until the transparent blue gel was seen. The simple model reaction for the creation of complex compounds may be expressed as followed (M=metal cation, $x$=1,2,3):

$$M^{2+} + nC_6H_{8-x}O_7^{x-} \leftrightarrow M(C_6H_7O_7)_n^{(xn-2)-}$$

It is essential for this reaction to keep the pH around 3.5 to avoid the possible precipitation; for this pH the CA dissolves with $x$ = 1÷2. After drying for a day in open air, the gel was subjected for the DTA/TG analysis (on TA SDT 2960 system) and on the basis of this the dried gel was first put onto the calcination at $550^0$C before the sintering. The purpose of such pre-sintering process is to burn off the remaining organic compounds. The dark powder, called the xerogel, was obtained at this stage. This powder was pressed into the cylinders of 10*mm* diameter and 1*mm* height and was then sintered at $900^0$C in air for 24h to produce the final $Ca_2CuO_3:U_x$ samples.

The SEM and EDX analysis were performed on the Jeol 5410LV and the Oxford ISIS 300 system. For the resistivity measurement, the gold contact was sputtered on both faces of the samples using the Leybold DC/RF Univex 450 system. The x-ray diffactograms were taken on the Bruker D5005 diffractometer with CuK$\alpha$ radiation, 2$\theta$ scan from 20-$70^0$, step angle $0.01^0$. The Raman scattering measurements were performed on the Dilor LABRAM 1B system equiped with He-Ne laser source ($\lambda$=632.8*nm*). The IR transmission spectra were obtained from the Fourier spectrometer with the CsI pellets having 0.5% of the finely ground samples.

**III. Structural characterization**

Figure 2 shows the DTA/TG analysis of the dried gels for two samples with $x = 0.0$ and 0.025 (other are omitted for clarity) where two peaks are clearly seen: the exothermal peak at $470^0$C and the endothermal peak with large weight loss at $810^0$C. The first one corresponds to the oxidation of the non-reacting citric acid and the organic middle products and the second one to the forming of the $Ca_2CuO_3$ phase. This temperature ($810^0$C) agrees quite well with the one reported in Huynh *et al* [23] and in Dou *et al* [1]. The weight loss at this temperature was caused by the freeing of the excessive oxygen.

Figure 3 shows the EDX analysis for the undoped sample (*x*=0.0) and one U-doped sample (*x*=0.05). For the undoped sample no other element except Ca, Cu and O was found. For the doped samples with $x < 0.05$ the presence of uranium was not clearly resolved and for $x = 0.05$ two small peaks for uranium appeared at 3.17 and 3.34 keV. In other studies of the uranium doped compounds, *e.g.* in the Tl-1223 superconductors, the uranium related peaks around 3.2 and 13.7 keV were also not seen for the molar ratio U/Ca < 0.1 [20].

In Figure 4(a) the SEM images of surfaces for the samples are shown. All samples appear to have the good phase homogeneity, smooth surfaces, but the greater grains and the more melted boundary are seen for the uranium free sample. The estimated average grain size is 640 *nm* for the uranium free sample and 380 *nm* for *x*=0.025. This refinement of the nanoparticle size due to the uranium doping has also been reported for the Nd-Ba-Cu-O system [17,18]. The reduction of grain size was accompanied by the fact that all doped samples appeared mechanically less fragile. The estimation of monocrystal size by the fourier analysis of peak shape for the strongest peak in the x-ray diffraction profile (using the WinFit software, see Ref. [35] for description of the method envolved) also showed an impact of the uranium doping on reducing the average size from 29.8*nm* (uranium free) to 22.1*nm* (*x*=0.025). So each polycrystalline nanoparticle (grain) in the pure $Ca_2CuO_3$ should contain approximately 21 monocrystals while in the doped samples it should contain only 17. This observation agrees quite well with the clearer rod-shape of grains for $x > 0$ (*i.e.* the preferred orientation is still well preserved along *c*-axis). Figure 4(b) shows the overall reduction of both nanoparticle size and monocrystal size due to the increased uranium content. The linearity is however not well defined and the effect seems to approach a limit.

To derive the crystal structure for the doped samples, the Rietveld analysis of the profiles was taken on the basis of the diffractograms shown in Figure 5 using the WinMProf software [34]. The final structural data are collected in Table 1 together with the R-factors. For all cases, the refinement was carried out using the same procedure and tactics as in Huynh *et al* [23]. First, we refined the profile using the pseudo-Voigt function including the correction on zero-point, background, asymmetry, half-width, scale factor and the cell parameters, then we refined the atomic coordinates, temperature factors, site occupation factors. The constraints were the same as in Ref. [23]. For the convenience of the readers in doing comparison of these results with the other already reported, the diffractograms in Figure 5 were indexed in the (*−cba*) axis orientation (same as in Ref. [23]) rather than in the (*abc*) orientation which was used in Lines *et al* [29] and in JCPDS files [33] (the transformation matrix [(0,0,1; 0,1,0; -1,0,0)] swaps the two). It is obvious from the given diffractograms that the purity of phases reached high grade, there is no trace for any of the following impure phases: $CaCu_2O_3$, CuO, CaO [33]. For the samples prepared by the other techniques, the impure phases were usually: CaO, CuO for the ceramic route (see Zhang *et al* [30] and discussion in Huynh *et al* [23]), CuO for the oxalate co-precipitation (Dou *et al* [1]), $CaCu_2O_3$ for the travelling-solvent floating-zone technique (Wada *et al* [24]). For the sol-gel route, it is worth considering the possible $UO_3$ phase since this phase is formed at lower temperature than the main phase ($500^0$C *vs.* $810^0$C). From the diffractograms given (Figure 5), the sol-gel method seemed to successfully exclude this impure phase as it was not observed in the whole 2θ scale.

The substitution of uranium did not induce any systematical change into the lattice parameters of the doped samples. While the cell constant *b* stayed almost unchanged (3.778Å), the *a* and *c* fluctuated randomly. All these cell constants agree very well with the ones reported for the single-phased $Ca_2CuO_3$ [1, 23, 30, 33]. In agreement with the fluctuation of *a* and *c*, the vertical bonds Cu-O$(1)_2$ and Ca-O$(1)_1$ show larger variation than the horizontal bonds Cu-O$(2)_2$ and Ca-O$(1)_4$. The rigidity of the Cu-O$(2)_2$ bonds (*i.e.* of the *b*-axis) and the elasticity of the Cu-O$(1)_2$ and Ca-O$(1)_2$ bonds (*i.e.* the relaxing position of Ca and O(1) along *c*-axis) appeared as the characteristic features for the structures under consideration. In the space group *Immm* the only positions that are allowed to vary are the *z*-coordinates for the Ca and O(1). Huynh *et al* [23] showed that this relaxation along axis *c* provides an intuitive way to interpret the strong appearance of two Raman-allowed $A_g$-mode phonons observed at 306 and 530 $cm^{-1}$ in the Sr-doped and pure $Ca_2CuO_3$ [14, 38, 39].

The small and random variation of the *a, c*-axis here differs from the situation reported for the Tl-1223 system [20] and the PZT (Mg/Nb) ceramics [21], where the systematical shortening of the *c*-axis was seen (at the uranium concentration 0.01/Ca the shortening was 0.05Å [20]). As suggested in Ref. [20], this shortening may indicate the incorporation of uranium into the parent lattice, specifically into the $Ca^{2+}$ sites. This suggestion is worths a careful consideration. Despite the variety of oxidation states, uranium usually occured in the doped ceramics in the stable oxidation state $U^{6+}$ and due to small ionic radius (0.87Å) it fits quite well into various lattice positions. Since in the $Ca_2CuO_3$ structure, the $Ca^{2+}$ cations (having radius 1.14Å) lie along axis *c*, the substitution of the smaller $U^{6+}$ cation would shorten the *c*-axis in compatible measure. Concretely, for $x$=0.025 each 25 among 1000 unit cells will have the *c*-axis shortened about 0.27Å, producing an average contraction around 0.007Å. A support argument for this comes from the statistical crystallography of the so-called Voronoi-Dirichlet polyhedra (VDP) of uranium [25]. By definition the VDP of an atom A surrounded by atoms $X_i$ is a convex polyhedron where each surface passes normally through the middle point of the corresponding $A–X_i$ bond. The calculation of VDP of the uranium substituted for $Ca^{2+}$ in the $Ca_2CuO_3$ yields 14.3Å$^3$ which is 55% greater than the average 9.2(3)Å$^3$ obtained from 354 independent structures containing $U^{6+}$ cation [25]. So the lattice parameters of the cell containing U should contract, specifically 0.2Å along the *c*-axis.

This scenario, however, suffers from the non-stoichiometry of the compound in which the $Ca^{2+}$ cation is replaced by the $U^{6+}$. With such replacement, two additional oxygens would be needed to preserve the stoichiometry: one should add a conner to a pyramid around $U^{6+}$ and transforms it into an octaheder (in all 4 polymorphs α, β, γ, δ of $UO_3$ the preferred coordination of the uranium is octaheder) and another should compensate the lack of the neighbour $Ca^{2+}$ covalence by the same creation of the octaheder. This would break down the regularity in the $Ca_2CuO_3$ lattice and create the dot defect. Further effect should be considered: the freeing $Ca^{2+}$ cations would recombine with the oxygens to form CaO. At high doping concentration this impurity might be substantial. In case the uranium would subsitute for both $Ca^{2+}$ and $Cu^{2+}$, the impure phases became both CaO, CuO and its possible product $CaCu_2O_3$. So the systematical incorporation of uranium into the parent lattice would not be without any damage to this lattice. Lines *et al* [29] showed that the cation oxidation states in the $Ca_2CuO_3$ were exactly 2+ (see also Ref. [23]) and the oxygen-excessive compound $Ca_2CuO_{3+\delta}$ could not be prepared under oxygen pressure at 400 *at* (the pressure at which the $K_2NiF_4$ type structure superconducting $Sr_2CuO_{3+\delta}$ was prepared). This means that for forming of the ideal $Ca_2CuO_3$ lattice the matching of exact stoichiometry is a crucial factor. For this reason, if the uranium was incorporated into the parent lattice, it would preferentially occur at the interstitial regions between the monocrystal pieces and represent themselves as the dot defects. Consequently, the uranium should not introduce change into the lattice parameters but should affect the number of defects. With increased doping concentration, this number increases and it is reasonable to assume that this effect either limits or reduces the size of both monocrystals and nanoparticles (Figure 4(b)). For the other known cases, the occurence of uranium in the parent structures can only be partly identified [26, 27].

## IV. Optical phonons

The incorporation of uranium into the parent structure leads to some speculation upon its effect on the Raman spectra of the doped samples. For the pure and Sr-doped $Ca_2CuO_3$, several Raman studies are available [14, 38, 39]. According to the optical conductivity measurement [40], the maximal scattering output is expected near the 2.0 eV region and we conducted the Raman scattering measurement using the light from He-Ne laser with $\lambda$ = 623.8 *nm* (1.96 eV). The same laser has been used in Yoshida *et al* [14]. The results are showed in Figure 6, from which peaks are seen for $x$=0.0 at 200, 280, 307, 467, 530, 663, 890, 942, 1142, 1217, 1337 $cm^{-1}$. For the purpose of comparision (mainly with Ref. [38]), we have also recorded the Raman spectra using the Nd:YAG laser with $\lambda$=1064*nm* (1.17eV)).

Table 2 summaries all observed frequencies. Let us briefly discuss how these peaks have been assigned in Ref. [14], [39]. From the symmetry analysis, in the space group *Immm* ($D_{2h}^{25}$), the optical phonons at the $\Gamma$ point ($\boldsymbol{k}$ = **0**) are composed of 6 Raman active modes ($2A_g$+$2B_{1g}$+$2B_{2g}$) and 9 IR active modes ($3B_{1u}$+$3B_{2u}$+$3B_{3u}$). The $A_g$-, $B_{1g}$-, $B_{2g}$-mode phonons associate with the Wyckoff site *4f* (site symmetry $C_{2v}^{2}$) of the Ca and O(1), so with the vibrations of these atoms along axis *c* ($A_g$) and *a, b* ($B_{1g}$, $B_{2g}$).

The $A_g$-mode phonons are active in the (*a, a*), (*b, b*) and (*c, c*) geometry and the $B_{1g}$-, $B_{2g}$-mode phonons are allowed only in the (*a, c*) and (*b, c*) settings. So by performing the scattering measurement in these exact configurations with some single crystal samples, the $A_g$-, $B_{1g}$-, $B_{2g}$-mode phonons can be determined. Indeed, Yoshida *et al* [14] has identified the $A_g$-mode phonons to be 306 $cm^{-1}$ (assigned to the Ca movement) and 530 $cm^{-1}$ (assigned to the O(1) movement). These two phonons were the sole phonons in the $c(a,a)\bar{c}$ and $a(c,c)\bar{a}$ configurations, so the assignments were unique. Both of them agree well with our lines at 307 and 530 $cm^{-1}$. However, no structures due to $B_{1g}$-, $B_{2g}$-mode phonons were observed in the respective scattering configurations [14, 39].

The rich features only appeared for the $a(b,b)\bar{a}$ configuration, *i.e.* when the light polarization was parallel to axis *b*. Yoshida *et al.* [14] has reported the following lines: 235, 306, 440, 500, 690, 880, 940, 1140, 1200 and 1330 $cm^{-1}$. All these peaks, except the one at 500 $cm^{-1}$ (also not seen in Ref. [38], [39]), have their counterparts in our spectra. The weak features that were also visible in Ref. [14] (but not discussed) correspond closely to 200, 470, 640, 1000 and 1390 $cm^{-1}$; the first two of them were also reported in Ref. [39].

The $B_{1u}$-, $B_{2u}$-, $B_{3u}$-mode phonons, associated with all Wyckoff sites in the *Immm* space group (namely, *2d* of Cu, *2a* of O(2), *4f* of Ca and O(1)), correspond to the vibration of these atoms along the crystallographic axis *c, b* and *a* respectively. As these modes are IR active, they can be observed in the reflectivity measurement for light polarization along each axis [14] or in the IR transmission measurement (Ref. [39], *this* *work*). The following lines were reported in Ref. [14] (TO(LO)- phonons): 215(260), 340(430), 660(700) $cm^{-1}$ ($B_{2u}$); 260(290), 410(430), 460(480), 580(630) $cm^{-1}$ ($B_{1u}$ and $B_{3u}$); the additional structures were found at 350 and 540 $cm^{-1}$ and were ascribed as $B_{1u}$- or $B_{3u}$-mode phonons in Ref. [39]. Most of these peaks were reproduced in this work and in Ref. [39], where the overall agreement with calculated values was seen.

The Raman spectra of the $Ca_2CuO_3$ however showed much more richer structures than that offered by the symmetry analysis. The series of the Raman forbidden peaks was seen for both doped and undoped $Ca_2CuO_3$ (*this work,* Ref. [14], [38]). Among these peaks, the 440, 500 and 690 $cm^{-1}$ were ascribed as the first-order zone-boundary phonons (T-point with **k** = (0.5,0.5,0)), whereas the 880, 940, 1140, 1200 and 1330 $cm^{-1}$ as their high-order two phonon scattering [14]. Since the 440, 690 $cm^{-1}$ lines were also observed for the undoped samples (440, 670 $cm^{-1}$ in Ref. [38]; 430, 690 $cm^{-1}$ in Ref. [39]; 430, 670 in this work, $x = 0$) Zlateva *et al* [39] has suggested that all extra lines in the Raman spectra are due to the high-order scattering. This resources in the finite and segmented Cu-O(2) chains of different lengths, which leads expectably to the overtones. It may result, however, from the impure phases presented as it was difficult to exclude all CuO, CaO and $CaCu_2O_3$ phases from the final product by means of the ceramic and oxalate co-precipitation techniques [23, 1, 24, 29, 38].

For the purpose of classification of all vibrational states, we performed the *ab initio* study on a model cluster $Ca_{18}Cu_8O_{28}$ (Figure 1(b)) with the Gaussian 2003 software [36]. This is a medium size layer model stacking one Cu-O layer between the other two Ca-O layers. The calculation was performed using the Self-Consistent-Field (SCF) Hartree-Fock (HF) method with the unrestricted spin model (UHF) on the 3-21G wave function basic set. The task was accomplished with the Mulliken charge analysis and the thermochemistry analysis for the vibrational states. The details of this calculation is presented in [42], here we summarize the main issues.

From the analysis of the simulated vibrational states (Figure 8) three IR-active $B_{2u}$ frequencies 215, 350, 670 $cm^{-1}$ ($x$=0) correspond to the vibration of Ca, O(1) and O(2) along $b$ axis (these vibrations have been assigned in Ref. [14] to Cu, O(1) and O(2) respectively). The IR-peak at 194 $cm^{-1}$ observed in Ref. [39] is caused by the moving of O(2) along axis $a$. The ($B_{1u}$+$B_{3u}$)-phonons at 350 and 532 $cm^{-1}$ ($x$=0) associate with the vibration of the same atom along axis $a$ and $c$. The other ($B_{1u}$+$B_{3u}$)-phonons at 415 and 453 $cm^{-1}$ ($x=0$) originate with the moving of O(1) along $c$ and $a$ axis respectively. The rest peaks at 564-566 $cm^{-1}$ ($x$=0.1-0.5) are from the vibration of both O(1) and O(2) along axis $c$ and the peak at 272 $cm^{-1}$ ($x$=0) associates with the breathing type vibration of Ca. The assignments for the two Raman-active $A_g$-phonon 307 and 530 $cm^{-1}$ ($x$=0) are the same as in Ref. [14]. These phonons were caused by the moving of the atoms Ca and O(1) along axis $c$ in nearly the static host lattice.

Among the Raman-forbidden peaks that were considered as the overtones in the previous studies [14, 39], the peaks at 200, 240 and 280 $cm^{-1}$ ($x$=0, 0.025-0.05, 0) follow from the lattice breathing vibration (LBV) with the O(2) atoms moving mainly along axis $a$. The peaks 435 ($x$=0.025), 467 ($x$=0) and 500 $cm^{-1}$ (Ref. [14]) originate likely from the vibration of both O(1) and O(2) atoms. It is worth noting that the Raman shift at 500 $cm^{-1}$ (observed in Ref. [14] but not in our cases) may be due to the impure phase $Bi_2O_3$ that was present in the studied sample. Further, from the characteristic structure of peaks for the pure CuO phase (298, 345, 632$cm^{-1}$) one may expect that the shift at 640 $cm^{-1}$ observed in Ref. [14] probably points to the CuO impure phase. The peaks 235 and 1000 $cm^{-1}$ (Ref. [14]) (240, 1003-1005 $cm^{-1}$, *this work,* $x$=0.025 and 0.05) may also come from the impure CaO phase. From the valence charge analysis, the valence distributed within the Cu-O bonds in the pure CuO structure (6-coordinates with the average length 1.875Å) is a little higher than in the $Ca_2CuO_3$ (average length 1.889Å). This agrees with the smaller force constant for the Cu-O bonding in $Ca_2CuO_3$ which considerably reduced the associated Raman shifts to lower frequencies (*i.e.* 280-290 instead of 298 $cm^{-1}$).

For the Raman shifts which correspond to the vibration of Cu atoms, the *ab initio* results showed that there was no simple vibration of the Cu atoms in the static host lattice. All vibrations involving the Cu atoms are mainly the collective lattice-breathing type vibrations. This observation agrees well with the structural analysis on the rigidity of the Cu-O(2) bonds (axis *b*) which was previously given in Section III. The occurence of the forbidden lines along the $CuO_4$ chain direction (axis *b*) illustrates well the expected strong coupling of phonons in the $CuO_4$ chain with the electron-hole pairs created during excitation by light. Such coupling is a typical phenomenon in the superconducting cuprates.

Now let us consider the effect of the substitution of uranium on the behaviour of Raman spectra of the doped compounds. In the Sr-doped $Ca_{2-x}Sr_xCuO_3$ ($x$ = 0.2, 0.4) [14, 39], because the Sr is about twice heavier than the Ca the systematical reduction in phonon frequencies has been seen (Table 2). For our cases, the reduction would be even more radical if the $U^{6+}$ would really occur in the $Ca^{2+}$ sites: this element is almost six times heavier than Ca. However, no substantial reduction was observed in all uranium doped samples. The frequencies varied only slightly and the changes did not conform to any scheme. The variation was negligible for the $A_g$-mode frequencies as they held within 307±2 and 530±3 $cm^{-1}$, but was larger for the $B_{1u}$, $B_{2u}$, $B_{3u}$ phonons and for the overtones. Some lines appeared only for the higher doping concentration, *i.e.* the lines at 240, 435, 1003-1005, 1380-1390 $cm^{-1}$ were seen clearly only for $x$ = 0.025 and 0.05. This observation agrees with the analysis of possible uranium position in the parent $Ca_2CuO_3$ structure previously given in Section III. If the uranium could only occur in the interstitial regions between the perfect lattice segments, then it would have stronger effect on the combined scattering lines than on the first order lines. This also explains why the number of the overtones increased with the uranium doping concentration.

**V. Electrical properties**

We now turn to the transport behaviours of the samples. The measurement of resistivity was carried out using the standard four probes technique with the electronic equipment from Bio-Rad which was capable of detecting the *pico-amper* current. The closed-cycle He refrigerator was used to adjust the temperature in 0.1K precision. While the uranium doping had no considerable effect on either structure or Raman spectra of the doped samples, it strongly influenced their resistivity. The small doping of uranium lowered the resistivity of the samples by an order of magnitude. This effect was also observed in the PZT ceramics [21] and then used as a technique to achieve and maintain the resistivity of those ceramics within the required regions.

Figure 9 shows the obtained development of resistivities according to 1/T. The linearity demonstrates the evidence for thermally activated conduction mode based on the band-gap model ($\ln(\rho) \propto E_a/k_B T$). The fit according to this model gave considerably better figure-of-merits in comparison with the other models, *e.g.* the small polaron model ($\ln(\rho/T) \propto 1/T$) or the variable range hopping model ($1/T^{1/4}$ law). The percolative conduction regime ($1/T^n$ law), which recently appeared essential for some Ru-doped manganates and ruthenates perovskites [28, 31], gave comparably good results but we do not expect this model can provide a meaningful explanation due to the insulator characters of the U-doped $Ca_2CuO_3$. The band-gap conduction mode has also been demonstrated as the most probable conduction model in the U-doped (Mg/Nb) PZT ceramics [21]. However, the additional modification to this model should be taken into account to explain the phonon-assisted hopping of electrons between the uranium dopant sites.

We may observe that the values of the activation energy $E_a$ are quite small, they all hold below 0.1 eV and lower considerably with increasing uranium content. The previous studies have reported $E_a$ = 0.18 eV for the undoped $Ca_2CuO_3$ [15, 32]. The higher activation energy was also found for the (Mg/Nb) PZT ceramics, *e.g.* with 0.48 mol% U-doping the activation energy was 0.34 eV. The small activation energy in our cases contrasts with the large resistivities of the samples and manifests the state so-called the *covalent insulation* in these compounds. The inset in Figure 9 shows the dependence of $E_a$ on the uranium content, as seen, the linearity is relatively weak (correlation coefficient $R^2 \approx 0.92$). It is worth to note that the obtained values of $E_a$ closely coincide with the observed HOMO/LUMO gap 0.082 eV (*i.e.* the gap between the Highest Occupied Molecular Orbital and the Lowest Unoccupied Molecular Orbital) from the *ab initio* calculation for the undoped cluster $Ca_{18}Cu_8O_{28}$ (the HOMO/LUMO could not be constructed for the doped samples due to the absence of the suitable wave function basic set for uranium). If we would consider the HOMO as the insulation ground state then the LUMO would be the first state in the conduction band (it is reasonable to assume that, the localization of the bonding molecular orbitals is a key factor for the covalent insulation state since the energy separation between the insulation and conduction states is small). For the uranium doped ceramics, the uranium atoms were usually regarded as the electron traping centers in the conduction band, so the HOMO-LUMO excitation would assist the hopping of electrons among the traping centers if the excitation would itself be coupled with the phonons. Indeed, the phenomenological concept of the phonon assisted electron hopping conduction mode proved to be successful in describing the resistivity behaviours of the uranium doped PZT ceramics [21].

It has been suggested that electrons are hopping between two dopant uranium sites with the probability $p$ equal to $kv\exp(-\alpha R)$ ($v$ is the phonon frequency, $k$ the constants, $\alpha$ the hopping parameter and $R$ the trap center distance). A good approximation for $R$ is $R = az^{-1/3}$ where $a$ is the lattice constant and $z$ the dopant concentration. The amplitude $\sigma_0$ of the $dc$-conductivity $\sigma$ should be proportional to the hopping probability $p$ and so might be given as $\sigma_0 = kv\exp(-\alpha az^{-1/3})$. Incorporating this into the band-gap classical expression for the conductivity $\sigma = \sigma_0\exp(-E_a/k_BT)$ we obtain the following relation for the $dc$-resistivity:

$$\rho = 1/\sigma = (1/kv)\exp(\alpha az^{-1/3} + E_a/k_BT). \qquad (1)$$

The fitting of $\ln(\rho)$ against $1/T$ according to this equation would yield the same $E_a$ as the fitting with the classical band-gap equation. However, the fitting of $\ln(\rho)$ against $z^{-1/3}$ would allow the determination of the hopping parameter $\alpha$. For our cases, we have obtained $\alpha = 0.18Å^{-1}$ for T=300K (using the lattice constant $a$=3.25Å). The fit is shown in the inset in Figure 10; note that the obtained slope is independent of phonon frequency $v$ and depends only on the linear distance $R$ between the two dopant sites, *i.e.* on the lattice constant $a$. The value $0.18Å^{-1}$ is a bit smaller than $\alpha = 0.63Å^{-1}$ reported in Ref. [21] for the U-doped PZT ceramics at the same temperature. To understand what this difference means, observe Figure 10 where the dependences of the normalized 'phonon independent' probabilities [$p\exp(a)/v$] on the dopant concentration $z$ for the two cases are shown. At the doping concentration $z = 0.05$, the curve for $\alpha = 0.18Å^{-1}$ shows approximately 15 times greater hopping probability than the curve for $\alpha = 0.63Å^{-1}$. The ratio of the non-normalized probabilities $p$ at this point is even higher, approximately 150 (of course, if the phonon frequencies are considered similar in both cases). This effect is due purely to the differences in the linear distance between the uranium dopant sites $R$. For the doping concentration $z$ from 0.005 to 0.05 in $Ca_2CuO_3$:$U_z$, the $R$ reduces from 19.0 to 8.8Å; the smaller value is about 2.7$a$. The most important outcome from equation (1) follows from the fact that the final conductivity depends only on the concentration of the dopant, not on the material composition. So this provides an easy way to compare the transport efficiency of various uranium doped systems despite their differences in chemical composition.

## VI. Conclusion

The uranium 238 can be doped successfully into the low dimensional system $Ca_2CuO_3$ by a modified sol-gel route with the citric acid as the reagent in keeping the pH for the reacting solution near 3.5. While the x-ray and Raman/IR spectra did not show any systematical variation due to doping, the particle size was substantially refined and the resistivity was significantly reduced in strong correlation with doping content. For the 1D spin chain system this achievement is remarkable since it implies that the chain length could successfully be controlled by doping. Recall that, the long-range antiferromagnetic order in the spin ½ chain system depends usually on the chain length and its local distortion. The fundamental aspect of the $Ca_2CuO_3$, from which various structural and vibrational states can be deduced or understood, is the rigidity of the Cu-O(2) bonding (*i.e.* of the *b* axis) and the relaxation of the Cu-O(1) and/or Ca-O(1) bondings (*i.e.* of the *c* axis). From the analysis given, the uranium is likely to occur at the interstitial regions between the monocrystal surfaces and represent themselves as the dot defects which, reasonably, require the further studies to classify the optical and related properties of the doped compounds. We leave this for future consideration.

**Acknowledgement**

The authors are grateful for the financial supports from the Grant Projects No. QG-07-02, KHCB 5.029.06, B2006-01-31 and DTCB 405 506.

TABLES

Table 1. The structural parameters for the $Ca_2CuO_3$:$U_x$. The space group is *Immm* (no.71). The standard deviations are given in parentheses. For the Wyckoff symbols of atoms and their fixed coordinates see Lines *et al* [29]. The *z*-positions for the Ca and O(1) atoms correspond to the (−*cba*) axis orientation (they should be the *x*-coordinates in the standard (*abc*) orientation). $R_P$ stands for the figure-of-merit based on the profile.

| *x* | *a* [Å] | *b* [Å] | *c* [Å] | *z*(Ca) | *z*[O(1)] | Cu-O(1)$_2$ | Cu-O(2)$_2$ | Ca-O(1)$_1$ | Ca-O(1)$_4$ | $R_P$ |
|---|---|---|---|---|---|---|---|---|---|---|
| 0.000 | 12.235(4) | 3.778(2) | 3.254(3) | 0.350(5) | 0.161(6) | 1.964(8) | 1.889(0) | 2.317(6) | 2.497(1) | 8.2 |
| 0.005 | 12.236(7) | 3.777(1) | 3.256(3) | 0.350(5) | 0.160(4) | 1.963(7) | 1.889(0) | 2.319(3) | 2.496(2) | 8.4 |
| 0.010 | 12.236(5) | 3.777(1) | 3.255(4) | 0.349(6) | 0.160(7) | 1.959(5) | 1.888(1) | 2.323(7) | 2.496(2) | 9.3 |
| 0.025 | 12.237(6) | 3.777(2) | 3.258(2) | 0.349(4) | 0.159(6) | 1.953(4) | 1.888(0) | 2.320(9) | 2.497(2) | 10.1 |
| 0.050 | 12.236(5) | 3.778(2) | 3.257(2) | 0.350(1) | 0.160(3) | 1.961(8) | 1.889(1) | 2.315(9) | 2.497(4) | 10.7 |

Table 2. The Raman and IR frequencies ($cm^{-1}$) for $Ca_2CuO_3$:$U_x$. Comparisons are given to the Sr-doped $Ca_2CuO_3$ [14,39] and to the theoretical values obtained by the lattice dynamic calculation [39] and the tight-binding approach [41]. For the Raman-forbidden lines, the values presented in parentheses correspond to the additional features visible in Figure 4, Ref. [14] but not reported by its authors.

| Phonon assignment | | Dopant concentration *x* | | | | | | | | | |
|---|---|---|---|---|---|---|---|---|---|---|---|
| (LBV=lattice breathing vibration) | | 0.2 (Sr) | 0.0 | Theory (*x=0.0*) | | | 0.0 | 0.005 | 0.01 | 0.025 | 0.05 |
| [14, 39] | *This work* | Ref.[14] | [39] | [39] | [41] | *This work* | *This work* | | | | |
| $A_g$-mode phonons (Raman active) (*c* axis) | | | | | | | | | | | |
| Ca | Ca | 306 | 311 | 311 | | 306 | 307 | 307 | 305 | 307 | 306 |
| O(1) | O(1) | 530 | 531 | 531 | 530 | 528 | 530 | 529 | 527 | 532 | 532 |
| $B_{2u}$-mode phonons (IR-active) (*b*-axis) | | | | | | | | | | | |
| Cu | Ca, Cu? | 215 | 225 | 201 | | 210 | 215 | 215 | 212 | 214 | 210 |
| O(1) | O(1) | 340 | 354 | 371 | | 337 | 350 | 352 | 350 | 348 | 351 |
| O(2) | O(2) | 660 | 682 | 673 | 700 | 657 | 670 | 667 | 668 | 670 | 670 |
| $B_{1u}$- and $B_{3u}$-mode phonons (IR-active) (*c* and *a* axis) | | | | | | | | | | | |
| Cu ($B_{3u}$) | Cu,O(2) ($B_{3u}$) | --- | 194 | 155 | 135 | --- | --- | --- | --- | --- | --- |
| Cu ($B_{1u}$) | Cu,Ca ($B_{1u}$) | 260 | 278 | 291 | | 265 | 272 | 270 | 272 | 270 | 270 |
| O(1),O(2) ($B_{3u}$) | O(2) ($B_{3u}$) | 350 | 354 | 337 | | 351 | 350 | 352 | 350 | 348 | 351 |
| O(1) ($B_{1u}$) | O(1) ($B_{1u}$) | 410 | 412 | 400 | | 410 | 415 | 413 | 415 | 417 | 415 |
| O(2) ($B_{3u}$) | O(1) ($B_{3u}$) | 460 | 457 | 424 | 450 | 457 | 453 | 455 | 450 | 450 | 452 |
| O(2) ($B_{1u}$) | O(2) ($B_{1u}$) | 540 | 530 | --- | | 548 | 532 | 530 | 535 | 534 | 535 |
| O(2) ($B_{1u}$) | O(1),O(2)($B_{1u}$) | 580 | --- | 577 | | 589 | --- | --- | 564 | 560 | 566 |
| The Raman-forbidden lines (overtones) | | | | | | | | | | | |
| ? | LBV+O(2) | (200) | 203 | | | 211 | 200 | 200 | 195 | 202 | 201 |
| Cu | LBV+O(2) or CaO | 235 | --- | | | 231 | --- | --- | --- | 240 | 240 |
| ? | LBV+O(2) or CuO | --- | 310 | | | 288 | 280 | 280 | 278 | 285 | 282 |
| T-point O(2) | LBV+O(1) | 440 | 430 | | 419 | 440 | --- | --- | --- | 435 | 433 |
| 235+235 | O(1)+O(2) | (470) | 472 | | | 461 | 467 | 466 | 465 | 468 | 467 |
| O(1), O(2) | O(1)+O(2) or $Bi_2O_3$ | 500 | --- | | 505 | 512 | --- | --- | --- | --- | --- |
| ? | CuO? | (640) | --- | | | 630 | --- | --- | --- | --- | --- |
| O(1), O(2) | ? | 690 | 690 | | 670 | | 663 | 660 | 663 | 665 | 664 |
| 2 phonon | 440+440 | 880 | 880 | | | | 890 | 889 | 888 | 892 | 892 |
| 2 phonon | 440+500 | 940 | 940 | | | | 942 | 937 | 939 | 943 | 941 |
| 2 phonon | 500+500 or CaO? | (1000) | | | | | --- | --- | 1003 | 1005 | 1003 |
| 2 phonon | 440+690 | 1140 | | | | | 1142 | 1140 | 1139 | 1145 | 1141 |
| 2 phonon | 500+690 | 1200 | | | | | 1217 | 1220 | 1215 | 1225 | 1222 |
| 3 phonon | 440+440+440 | 1330 | | | | | 1337 | 1340 | 1333 | 1340 | 1335 |
| 2 phonon | 690+690 | (1390) | | | | | --- | --- | 1380 | 1390 | 1388 |

FIGURES AND CAPTIONS (in printed size)

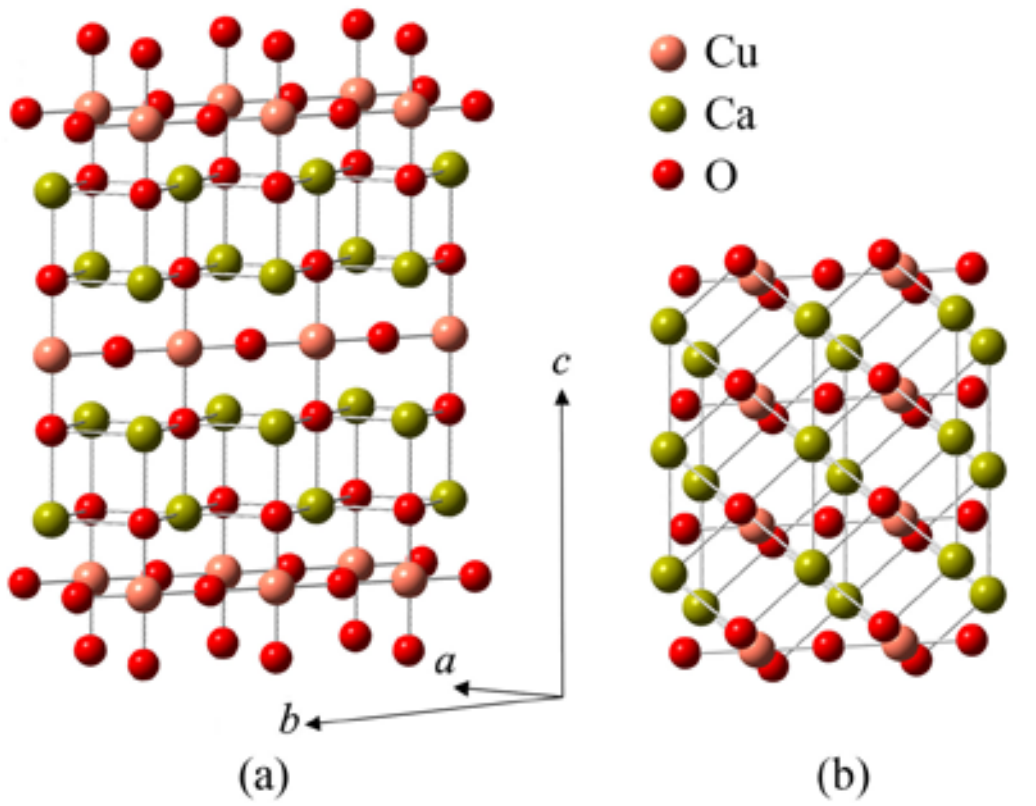

Figure 1. (Color online). The packing structure of 3 unit cells ($a\times 3b\times c$) for $Ca_2CuO_3$ (a) and the model cluster $Ca_{18}Cu_8O_{28}$ (b) used in the *ab initio* calculation of the vibrational states discussed in Section IV. The $CuO_4$ groups are connected through the O(2) atoms. The orientation of axes follows the (–*cba*) convention in which the *c*-axis is the longest. The coordination of Ca is the penta-oxygen bonding spheres $Ca^{2+}$–$O(1)_5$.

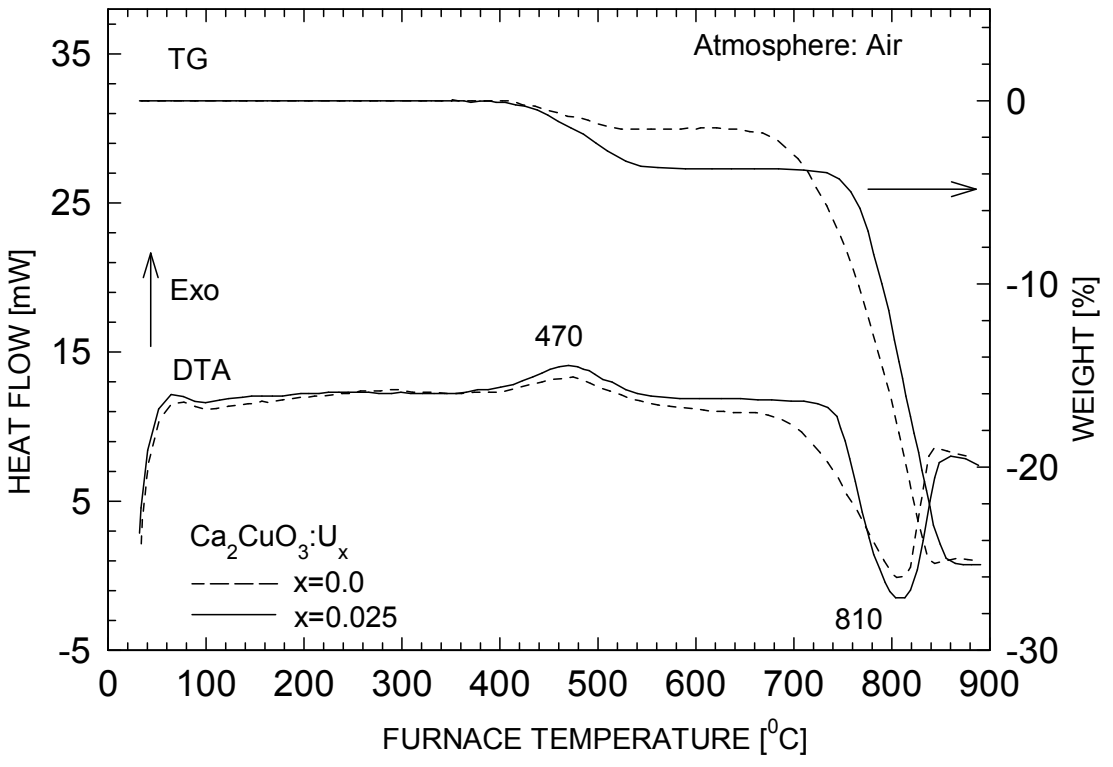

Figure 2. DTA/TG analysis of the dried gels for two compositions $x$ = 0.0 (dash line) and 0.025 (solid line).

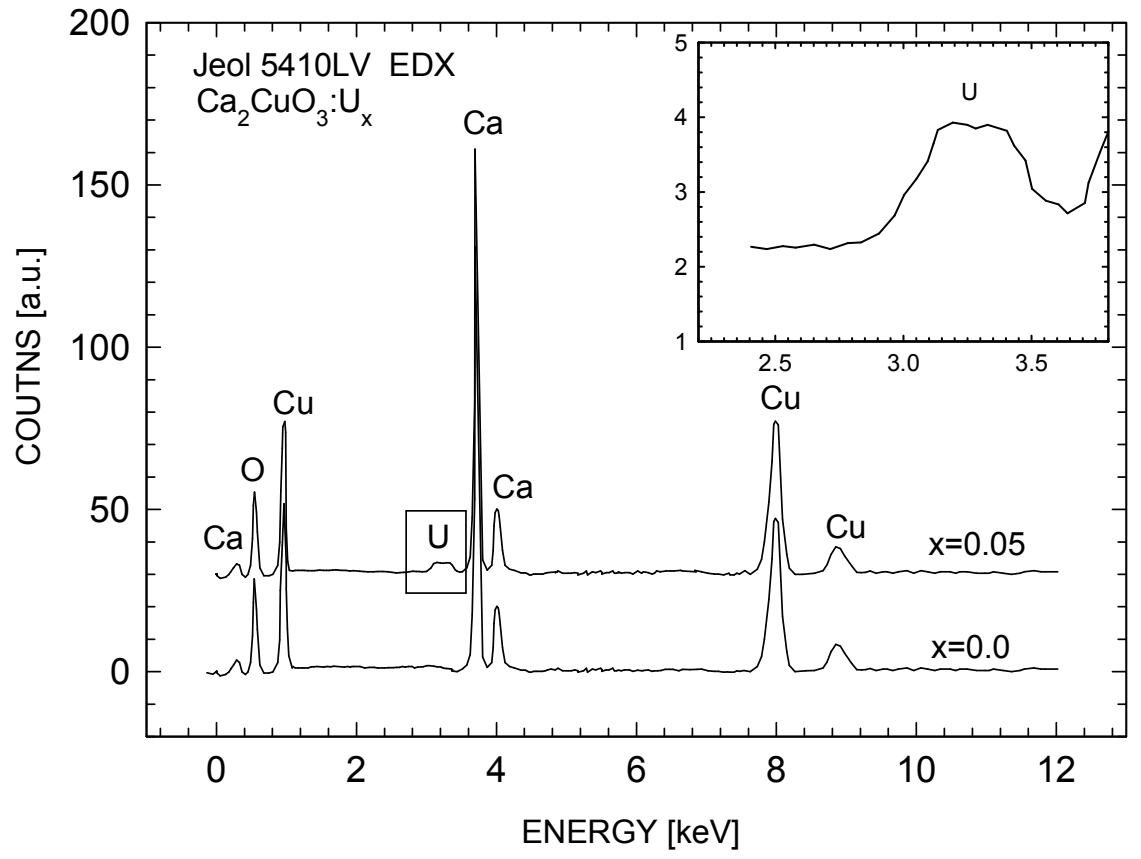

Figure 3. EDX spectra for the samples with $x$=0.0 and 0.05. The inset shows the zooming area near 3.2keV for the sample with $x$=0.05 which reveals two uranium related peaks at 3.17 and 3.34 keV.

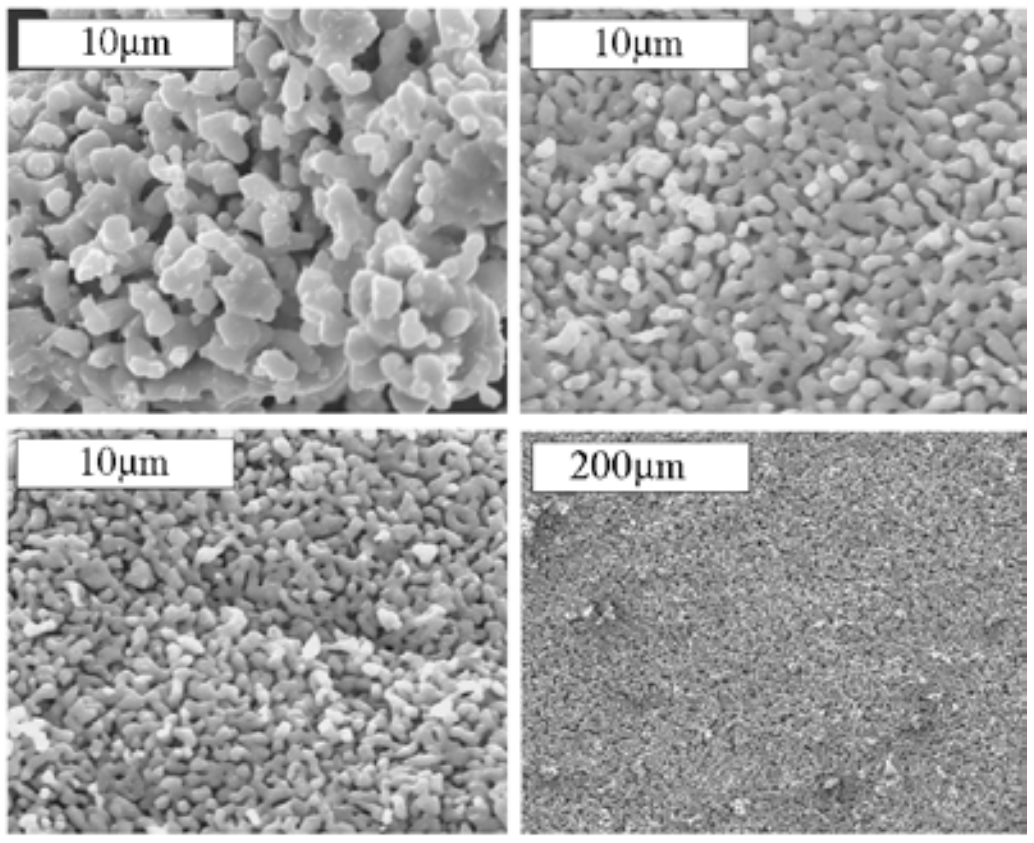

(a)

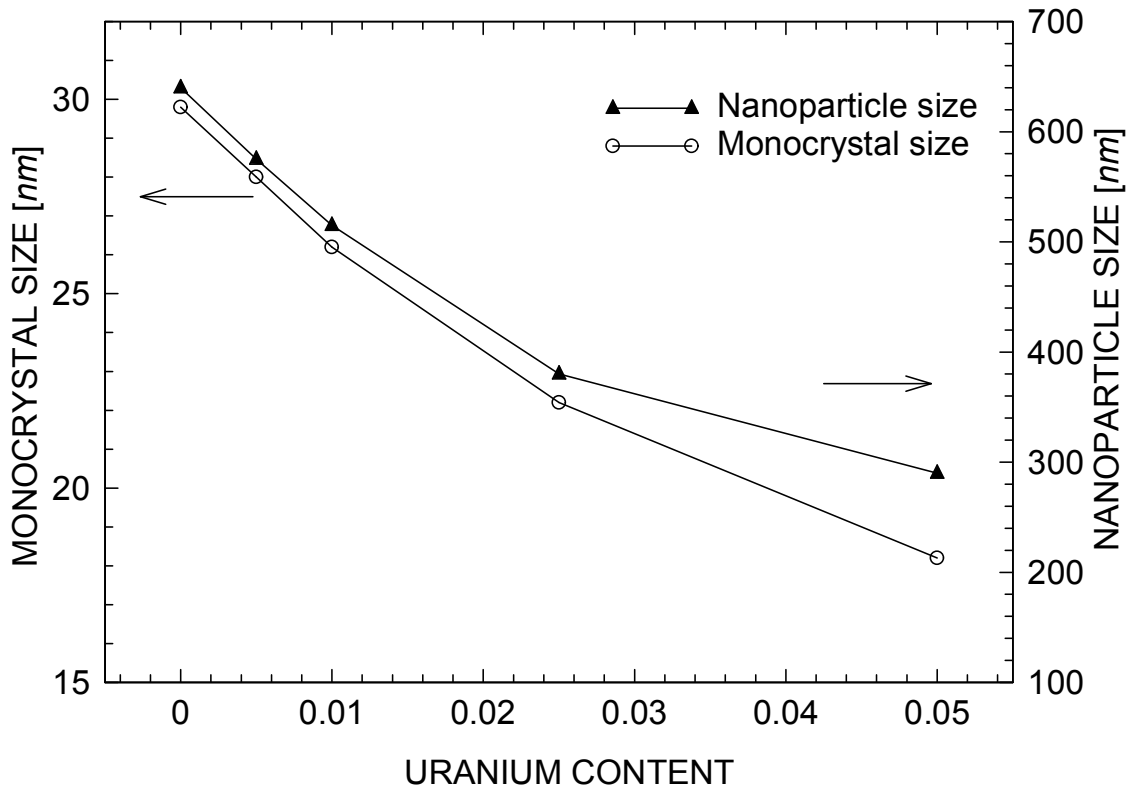

(b)

Figure 4. SEM photographs of the surfaces for the samples with $x = 0$, 0.1 (upper, left to right) and $x = 0.025$, 0.05 (lower, left to right) (a). The dependence of nanoparticle size and monocrystal size on uranium doping content (b).

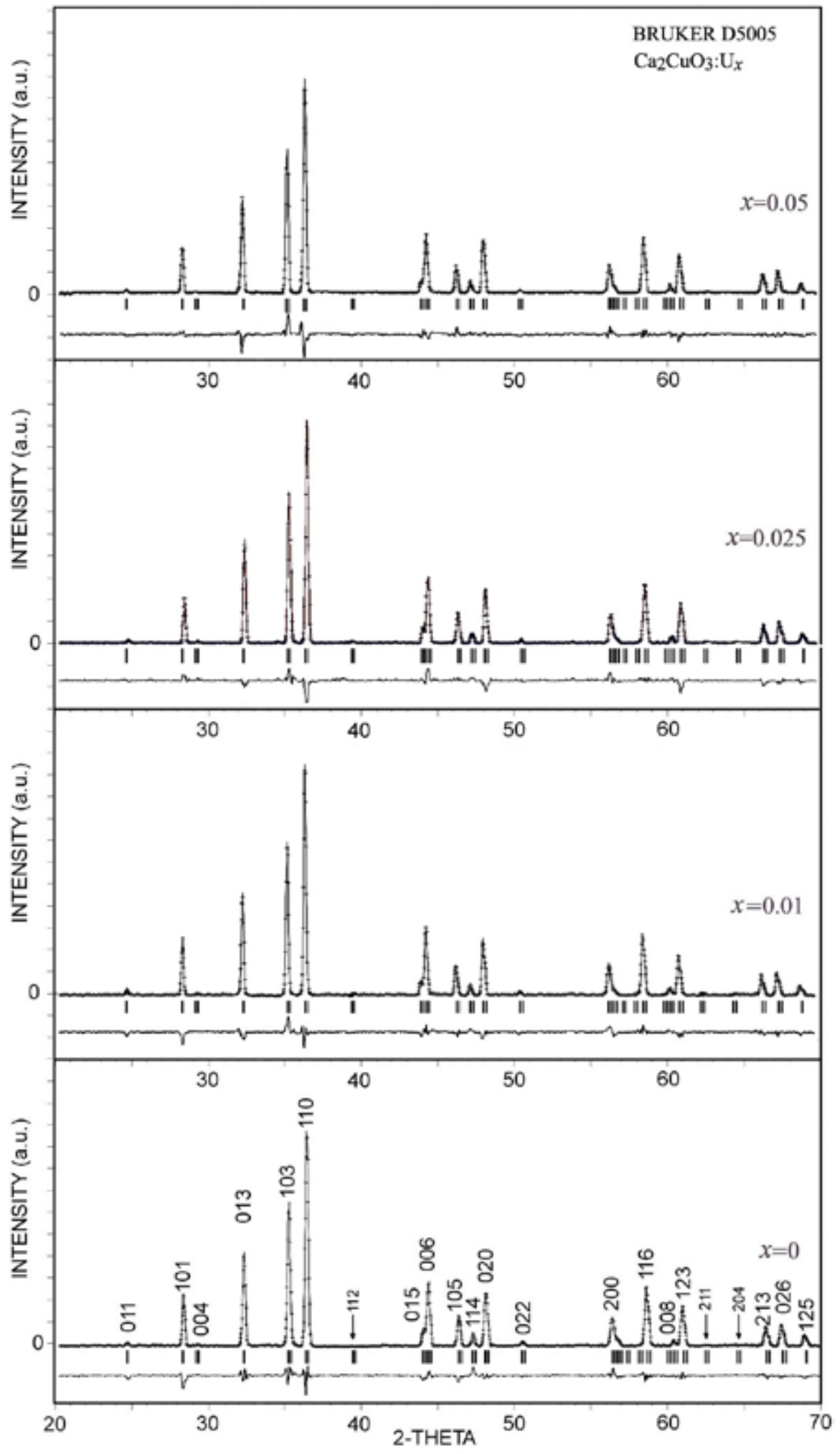

Figure 5. X-ray diffractograms with the profile fitting and different curves for the samples (the *x*=0.005 is omited since it is indifferent to *x*=0). The *hkl* indeces are given in the (−*cba*) axis orientation.

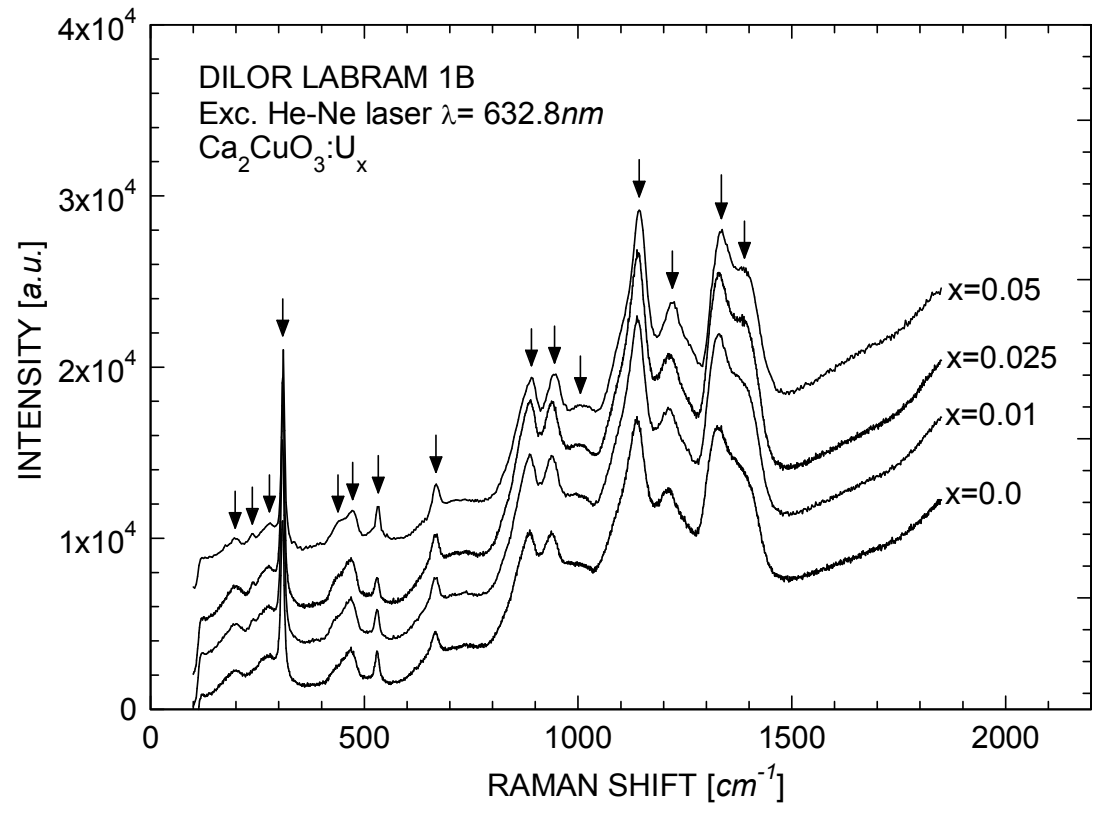

Figure 6. The Raman scattering spectra of the uranium doped $Ca_2CuO_3$ samples. The lines selected for listing in Table 2 are denoted by the arrows. The sample $x$=0.005 is omited since it is the same as $x$=0.

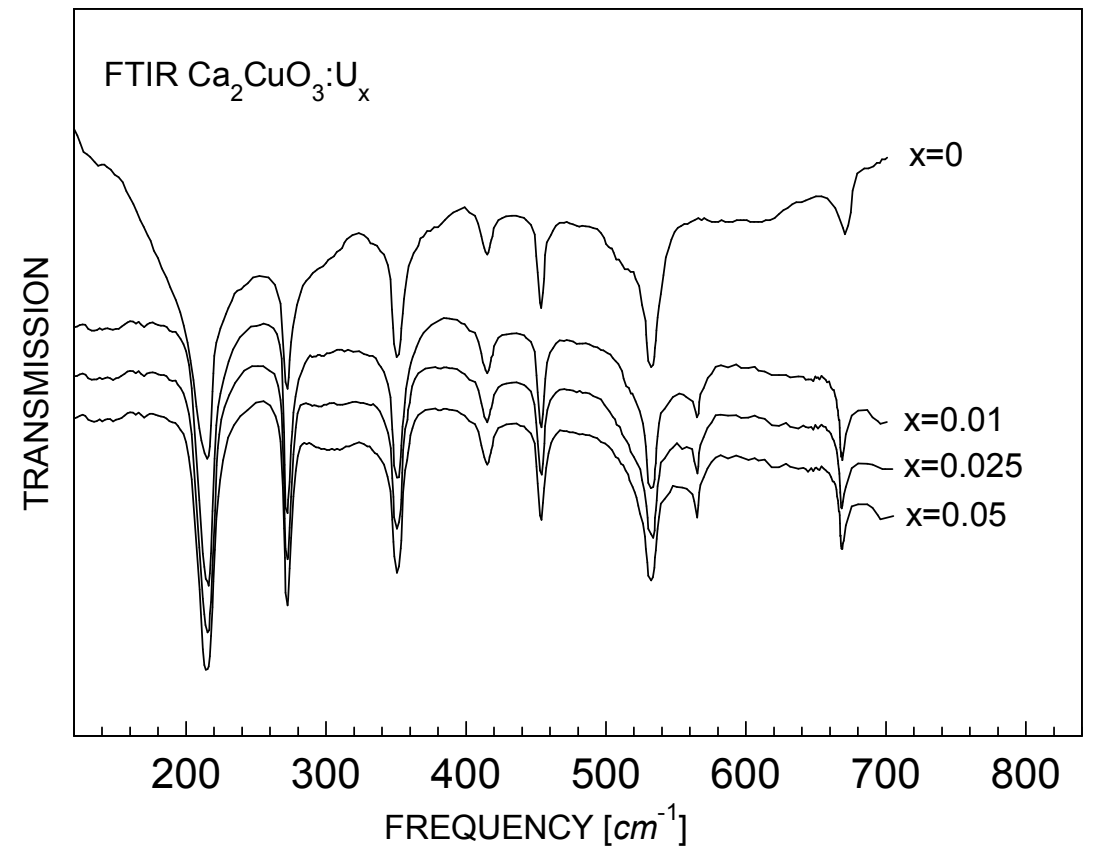

Figure 7. The IR transmission spectra for the uranium doped samples. The sample with $x$=0.005 is omited.

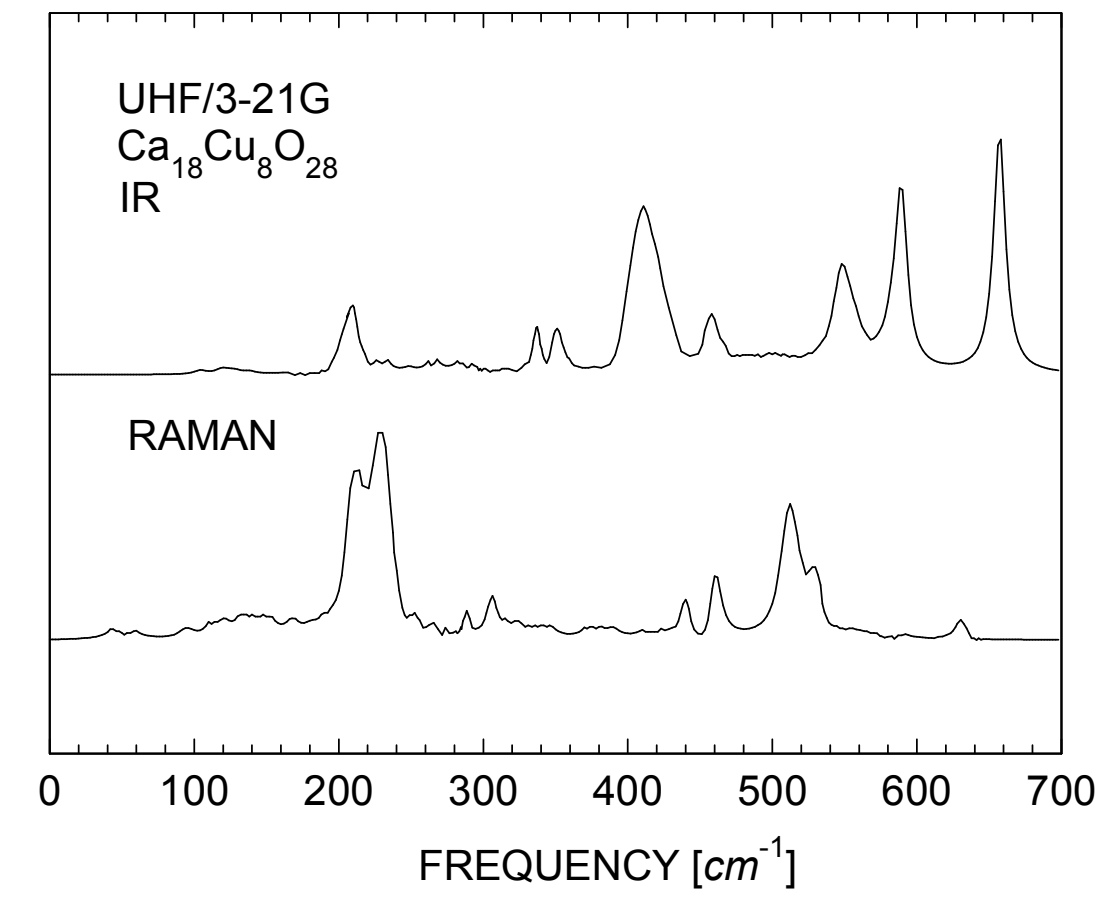

Figure 8. The simulated IR and Raman spectra for the $Ca_{18}Cu_8O_{28}$ cluster as obtained from the *ab initio* calculation using the Unrestricted spin Hartree-Fock SCF model with 3-21G basic set.

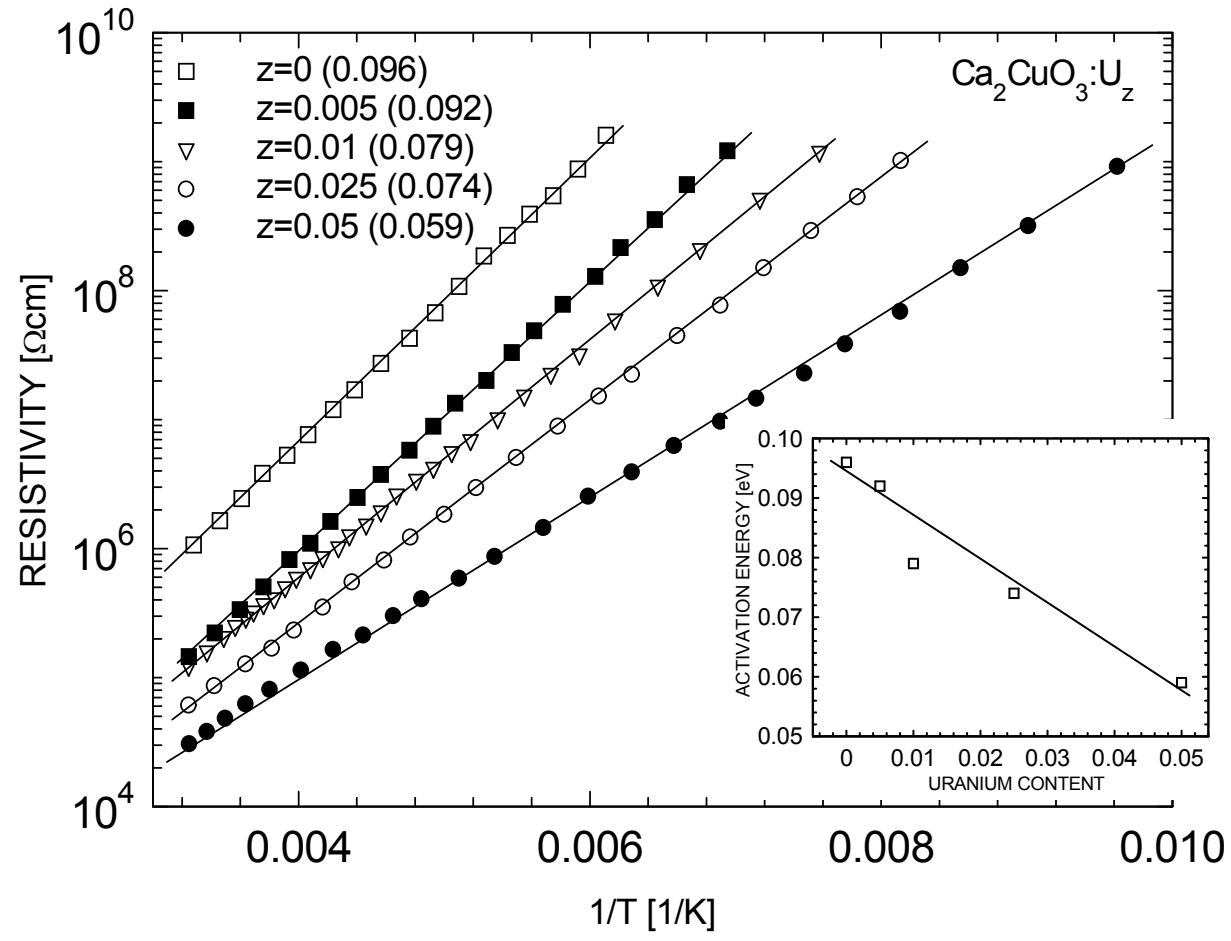

Figure 9. The resistivity of the samples in logarithmic scale drawn against 1/T. The values in parenthesis denote the activation energy $E_a$ as obtained from the slopes of the fitting straight lines. The inset shows the variation of this $E_a$ according to the uranium content.

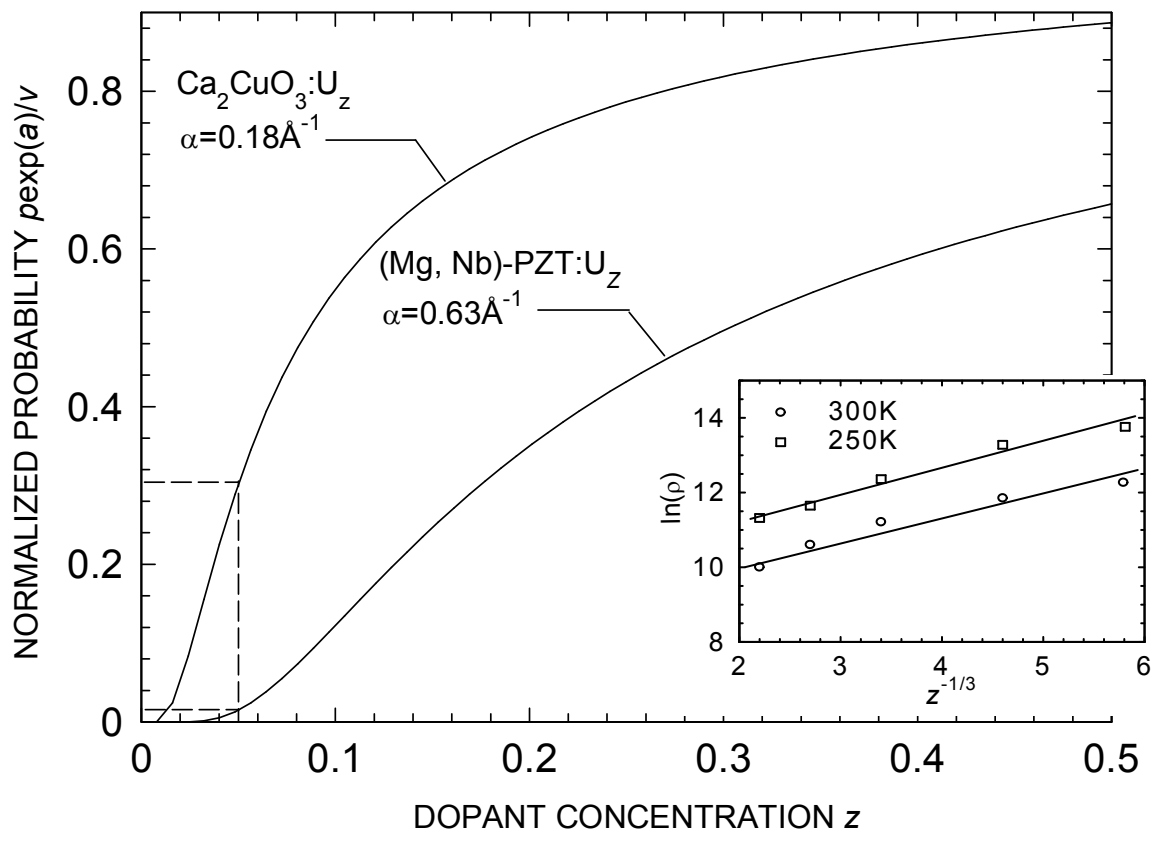

Figure 10. The normalized hopping probabilities [$p\exp(a)/v$] for two different uranium doped systems: the $Ca_2CuO_3$ with $\alpha$=0.18Å$^{-1}$ and the (Mg, Nb)-PZT ceramics with $\alpha$=0.63Å$^{-1}$ (Ref. [21]). The insets shows the fitting for $\alpha$ in two different temperatures 250 and 300K.